\documentclass[a4paper,11pt]{article}
\usepackage{pos}

\def\bea{\begin{eqnarray}}
\def\eea{\end{eqnarray}}
\def\nn{\nonumber}

\def\lmatrix{\left(\begin{array}}
\def\rmatrix{\end{array}\right)}

\def\msbar{\overline{\rm MS\kern-0.5pt}\kern0.5pt}
\def\W{W(F_4)}
\def\Wc{W(B_4)}

\title{Lattice QCD and the energy-momentum tensor on the 16-cell honeycomb}

\author{Sandor D. Katz}
\author*{Daniel Nogradi}

\affiliation{Eotvos Lorand University, Institute of Physics and Astronomy, 
    Department of Theoretical Physics\\
    Pazmany Peter setany 1/a, Budapest, 1117, Hungary}

\emailAdd{katz@bodri.elte.hu}
\emailAdd{nogradi@bodri.elte.hu}

\abstract{
    In previous work we formulated 
    lattice QCD with dynamical Wilson quarks on the 16-cell honeycomb lattice. The
    construction follows the usual cubic lattice formulation in that gauge fields are associated with
    links and fermions with sites, but the underlying lattice is different. The 16-cell honeycomb can be
    thought of as the union of two usual cubic lattices. Once one is given the other is shifted
    by half a lattice spacing
    in all directions, leading to 24 nearest neighbors instead of 8. An important consequence is a larger
    discrete space-time symmetry, leading to smaller cut-off effects in pure gauge and fermionic
    observables. The fate of doublers from the Wilson term is explained in detail. 
    The smaller space-time symmetry breaking relative to the cubic lattice also leads to more
    protection from operator mixing and less breaking up of continuum $O(4)$ irreducibles into 
    lattice symmetry irreducibles. As an example
    we show that the
    traceless energy-momentum tensor requires just two independent renormalization constants
    on the 16-cell honeycomb (same as in the continuum), and not four, as on the cubic lattice.
    The findings carry over to the 2-point function at non-zero space-time
    separation and the number of tensor structures with potential divergences, from the contact terms,
    is also reduced.
    }

\FullConference{The 43rd International Symposium on Lattice Field Theory (Lattice 2026)\\
July 26 to August 1, 2026\\
University of Maryland, College Park, USA\\}

\begin{document}
\maketitle

\section{Introduction}
\label{intro}

The 16-cell honeycomb lattice provides several advantages over the usual cubic lattice. Quantitatively,
the size of cut-off effects is much smaller in several gauge and fermionic observables
\cite{Katz:2025ycr}. Also, there is a qualitative difference in terms of operator mixing \cite{Katz:2026qmx}. 
These
latter results were obtained from the decomposition of the irreducible representations of the 
continuum Euclidean $O(4)$ space-time symmetry
group with respect to the subgroup which survives the lattice discretization.
The larger symmetry of the 16-cell honeycomb relative to the cubic lattice results in fewer 
factors in many decompositions, which leads to less independently renormalizing 
channels at finite lattice spacing, hence less mixing.

In the present contribution we review the details of the Wilson-Dirac operator construction and the fate
of doublers under the Wilson term. Furthermore, the observation about less mixing 
is applied to the energy momentum tensor. The corresponding
$O(4)$ representation for the traceless part is $H_2$ (in the notation of \cite{Katz:2026qmx}), the 9 dimensional
representation of 2-index symmetric traceless tensors. On the cubic lattice it breaks up into a 3 and a 6
dimensional irreducible representation of the symmetry group of the lattice, while on the 16-cell
honeycomb it remains a 9 dimensional irreducible. Thus for each of the two flavor singlet
dimension 4 operators, the gluonic and fermionic, there is a single renormalization constant for all 9
components, and not 2, as on the cubic lattice.

It should be noted that another approach, based on the small flow time expansion of the gradient flow 
\cite{Luscher:2010iy}
exists for continuum-like renormalization of the energy momentum tensor
\cite{Suzuki:2013gza,Makino:2014taa,Capponi:2015ucc,Capponi:2015ahp,DelDebbio:2020amx,Suzuki:2021tlr,Lee:2022mlh}.
In this approach however a well-known window problem enters, $a \ll \sqrt{8t} \ll 1/\Lambda_{QCD}$.
It amounts to ensuring the flow time is large enough so that
cut-off effects are small while at the same time the flow time is small enough in physical units so that
the small flow time expansion is applicable, which are difficult to fulfill in practice. 

For early work on lattice gauge theory and scalar QFT on the 16-cell honeycomb see
\cite{Celmaster:1982ht,Celmaster:1983jq,Celmaster:1983vy,Celmaster:1983hs,
Celmaster:1984rf,Celmaster:1985cv,Celmaster:1986eb,Neuberger:1987kt, Bhanot:1990zd, Bhanot:1990ai}.

\section{Wilson-Dirac fermions}
\label{doublers}

The construction of the Dirac operator is given in \cite{Katz:2025ycr} starting from the naive
discretization and adding a Wilson term. The naive discretization leads to a free propagator with a
number of poles, not only the physical one. 
There are a total of 72 poles and they come in groups with definite chiralities, subject to the sum
rule from the Nielsen-Ninomiya theorem. The multiplicity of each group is $(1,12,32,24,3)$ and the
chiralities are $(+,-,+,-,+)$, thus we have,
\bea
1 + 12 + 32 + 24 + 3 &=& 72 \nn \\
1 - 12 + 32 - 24 + 3 &=& 0.
\eea
The corresponding multiplicities and chiralities on the cubic lattice are the familiar,
\bea
1 + 4 + 6 + 4 + 1 &=& 16 \nn \\
1 - 4 + 6 - 4 + 1 &=& 0.
\eea
Let us call points in the Brillouin zone
where the propagator has a pole $p_0$, so there are 16 and 72 such 
points on the cubic
and 16-cell honeycomb lattices, respectively. Then one may expand around each, 
$p_0 + p$, for small $p$,
leading to a quadratic expression, which corresponds to the dispersion relation of the given particle
represented by the pole. 
In a Lorentz (or rather Euclidean) invariant theory this should be $p^2$. On both lattices for the
physical mode $p_0=0$ one indeed recovers $p^2$. On the cubic lattice this holds for all 15 other
doublers too. What was noted, however, already in \cite{Celmaster:1983jq} is that on the 16-cell honeycomb 
some of the doublers
give a quadratic expression which is not $p^2$, but rather $d_{\mu\nu} p_\mu p_\nu$ with $d$ not 
necessarily the identity. Some of them are proportional 
to the identity, some of them are not. These latter ones break
Lorentz (Euclidean) symmetry. The eigenvalues of the dispersion relation matrix $d$ are listed below for
each pole group,
\bea
1:&& (1,1,1,1) \nn \\
12:&& (\tfrac19,\tfrac19,\tfrac19,1) \nn \\
32:&& (\tfrac{1}{16},\tfrac{1}{16},\tfrac{1}{4},\tfrac{1}{4}) \\
24:&& (\tfrac19,\tfrac19,\tfrac19,\tfrac19) \nn \\
3:&& (\tfrac19,\tfrac19,\tfrac19,\tfrac19). \nn
\eea
Clearly, 44 doublers break Lorentz (Euclidean) symmetry, leading to the likely conclusion that the naive
discretization in the continuum limit is not Lorentz (Euclidean) invariant. The solution, given in
\cite{Katz:2025ycr} was to simply introduce a Wilson term, completely analogously to the cubic lattice,
which gives all 71 doublers a mass on the scale $1/a$ leaving behind just the physical mode, which as we
have seen, possesses a Lorentz (Euclidean) invariant dispersion relation.

\begin{figure}
\begin{center}
    \raisebox{0.75cm}{\includegraphics[width=6.4cm]{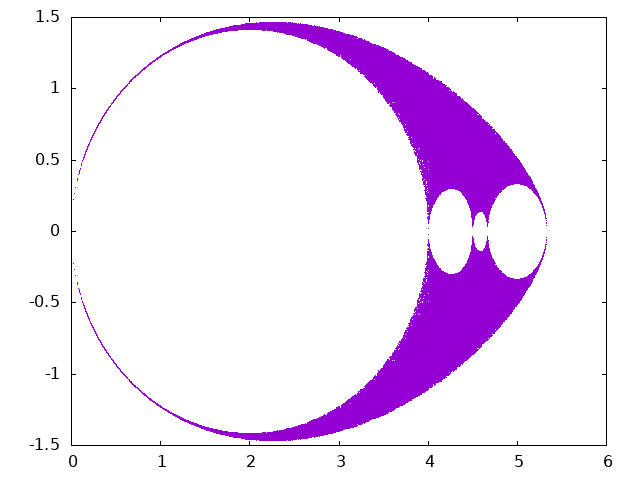}} \hspace{0.5cm} \includegraphics[width=8cm]{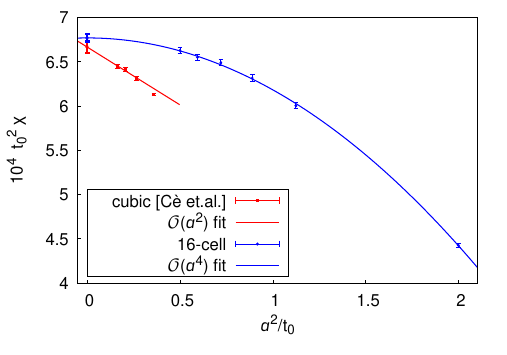}
    \caption{Left: free Wilson-Dirac spectrum in lattice units. Right: 
    Continuum limit of the topological susceptibility in pure $SU(3)$ gauge theory, in
    $t_0$-units on both the cubic lattice \cite{Ce:2015qha} and the 
    16-cell honeycomb \cite{Katz:2025ycr}.}
    \label{fig}
\end{center}
\end{figure}

More precisely, the Wilson term with Wilson parameter $r=1$ shifts the poles 
by the following amounts, in units of $1/a$, 
for each of the pole groups,
\bea
1: 0,\qquad\quad  12: 4,\qquad\quad 32: \frac92,\qquad\quad 24: \frac{14}{3},\qquad\quad 3: \frac{16}{3},
\eea
leading to the free Wilson-Dirac spectrum in the left panel of
figure \ref{fig}. The thinness of the spectrum, starting
from the physical mode at the origin 
all the way up to $\sim 2/a$ is striking, especially if compared with the
corresponding picture on the cubic lattice. This observation alone explains much of the reduction in
cut-off effects in fermionic quantities, such as (near) zero modes and their chirality, 
observed in \cite{Katz:2025ycr}. As an example of a pure gauge quantity, we show in the right panel of 
figure \ref{fig} 
the continuum limit of the topological susceptibility in $t_0$-units for both the cubic lattice 
\cite{Ce:2015qha} and the 16-cell honeycomb \cite{Katz:2025ycr}. The reduction in cut-off effects is
again striking.

\section{Energy-momentum tensor}
\label{emt}

Small cut-off effects are not only a quantitative observation, but in many cases lead to qualitative
changes, relative to the cubic lattice. What is relevant for the energy-momentum tensor is the
representation theory of the finite group of symmetries of the lattice. The cubic lattice has a symmetry
group with 384 elements, while the 16-cell honeycomb has a 3 times larger group. For details see
\cite{Katz:2026qmx}.

We use the group theoretic results from \cite{Katz:2026qmx}, the key ingredients can be deduced from the tables
there. Let us denote the symmetry group of the cubic lattice by $\Wc$ and the symmetry group of the
16-cell honeycomb by $\W$. The notation is explained in \cite{Katz:2026qmx}, it has to do with the Weyl group
of the two root systems, $B_4$ and $F_4$, respectively. We will use notation for the irreducible
representations of the two symmetry groups also from \cite{Katz:2026qmx}, namely $\tau^{(k)}_l$ will label the $k$
dimensional representations of $\Wc$ and $l$ labels the inequivalent ones for the same dimension, while
$\varrho^{(k)}_l$ is analogous for $\W$. There are a total of 20 irreducible representations of $\Wc$
while 25 of $\W$, and we have $|\Wc| = 384$ and $|\W| = 1152$.  

The traceless part of the symmetric energy-momentum tensor, $T_{\mu\nu}$, is in the 9 dimensional
representation of $O(4)$, denoted here by $H_2$. The scalar trace part is not considered.
The two independent gauge invariant flavor singlet operators in $H_2$, one gluonic and one fermionic, 
mix, leading to a $2\times 2$ mixing matrix. The anomalous dimension matrix has a zero
eigenvalue, corresponding precisely to the traceless energy-momentum tensor while another linear
combination runs logarithmically. Hence for the traceless energy-momentum tensor
two independent renormalization constants, $Z_g$ and $Z_f$, are sufficient.  

The different behavior on the cubic lattice and 16-cell honeycomb originates in the fact that $H_2$
is not irreducible under $\Wc$ but it is under $\W$. More precisely,
we have $H_2 = \tau^{(3)}_1 \oplus \tau^{(6)}_3$
and $H_2 = \varrho^{(9)}_1$. 
In the former case \cite{Caracciolo:1989pt,Caracciolo:1989bu,DallaBrida:2020gux}
the two components are the 3 diagonal, $\tau^{(3)}_1$, and the 6 off-diagonal,
$\tau^{(6)}_3$, entries.
Since the two sets renormalize independently, the number of renormalization constants is doubled to
four, $Z^{(3,6)}_{f,g}$,
and $Z^{(3,6)}_g$ are not the same at finite lattice spacing and similarly for $Z^{(3,6)}_f$.

On the 16-cell honeycomb $H_2$ stays a single irreducible one under $\W$. Hence the renormalization
properties are exactly the same as in the continuum, only two renormalization constants are necessary for
all 9 components.

Naturally, this observation carries over to the 2-point function of the energy-momentum tensor 
as well. Let us denote by $\Theta(x)$ the trace and by $T_{\mu\nu}(x)$ the traceless part, which in the
continuum is in $H_2$. 
Conservation connects the two by $\partial_\mu T_{\mu\nu} = - \tfrac14 \partial_\nu \Theta$.
In momentum space and in the continuum,
Ward identities limit the possible Lorentz structures and the 2-point functions can be written in terms
of just two scalars, the spin-2 and spin-0 components,
\bea
\langle T_{\mu\nu}(p) T_{\rho\sigma}(-p) \rangle &=& \
\frac{4}{3}\, Q_{\mu\nu} Q_{\rho\sigma}\, G_0(p^2) + 
\left( \frac12\left( P_{\mu\rho} P_{\nu\sigma} + P_{\mu\sigma} P_{\nu\rho} \right)  - \frac13 P_{\mu\nu}
P_{\rho\sigma} \right) \, G_2(p^2) \nn \\
\langle T_{\mu\nu}(p)\Theta(-p)\rangle &=& -4\, Q_{\mu\nu} G_0(p^2) \\
\langle \Theta(p) \Theta(-p)\rangle &=& 12\, G_0(p^2), \nn
\eea
understood up to contact terms, where we have,
\bea
P_{\mu\nu} &=& \delta_{\mu\nu} - \frac{p_\mu p_\nu}{p^2} \\
Q_{\mu\nu} &=& \frac{p_\mu p_\nu}{p^2} - \frac14 \delta_{\mu\nu}. \nn
\eea
Renormalization, in real space, at $x \neq 0$ so up to contact terms,
is fully determined by the same renormalization constants we have found above. 

On the lattice, contact terms bring additional additive divergences in the 2-point function and the
difference in symmetry groups between the cubic and 16-cell honeycomb 
leads to different number of allowed tensors. Although in many applications contact terms are eliminated by
a single suitable subtraction (for example finite temperature and zero temperature) 
it is instructive to see how the allowed number of tensors is reduced on the 16-cell honeycomb,
relative to the cubic lattice.

The contact terms may carry $a^{-4}$ and $a^{-2}$ power divergences. 
Let us consider the tensor structures at each order, first in the continuum. 
The independent number of tensor
structures follow from the fact that the energy-momentum tensor 2-point function lies in $Sym^2(H_2) = 
1 \oplus H_2 \oplus H_4 \oplus W$, where $H_4$ is the space of 4-index traceless symmetric tensors 
and $W$ is also 4-index and traceless but with the symmetry and anti-symmetry
of the curvature tensor of general relativity (Weyl tensor).
The tensors corresponding to $a^{-4}$ are $p$-independent so the single trivial representation component
gives a single tensor. At $a^{-2}$ we need a 6-index tensor contracted with the symmetric $p_\alpha
p_\beta$, which is in $1 \oplus H_2$. Then we see that $Sym^2(H_2) \otimes ( 1 \oplus H_2 )$ has two trivial
components, one from $1 \otimes 1$ and another one from $H_2 \otimes H_2$. The former corresponds to
simply $p^2$ times the $a^{-4}$ tensor, and the latter is a new independent one. Concretely, 
\bea
a^{-4}:&& \quad C^0_{\mu\nu\rho\sigma} = \frac12 \left( \delta_{\mu\rho} \delta_{\nu\sigma} + \delta_{\mu\sigma} \delta_{\nu\rho} \right) -
\frac14 \delta_{\mu\nu} \delta_{\rho\sigma} \\
a^{-2}:&& \quad C^{21}_{\mu\nu\rho\sigma} =  p^2 C^0_{\mu\nu\rho\sigma} \nn \\
&& \quad C^{22}_{\mu\nu\rho\sigma} = 
\frac{p^2}{2} \left( 
Q_{\mu\rho} \delta_{\nu\sigma} +  
Q_{\nu\rho} \delta_{\mu\sigma} +  
Q_{\mu\sigma} \delta_{\nu\rho} +  
Q_{\nu\sigma} \delta_{\mu\rho} 
\right) - \frac{p^2}{2} \left( Q_{\mu\nu} \delta_{\rho\sigma} + Q_{\rho\sigma} \delta_{\mu\nu} \right)  \nn 
\eea
On the cubic lattice at $a^{-4}$ we have, $H_2 = \tau^{(3)}_1 \oplus
\tau^{(6)}_3$ so it decomposes into two components, leading to two trivial representations in $Sym^2(H_2)$,
which is already one more than in the continuum. On the 16-cell honeycomb $H_2 = \varrho^{(9)}_1$, a single
component, so $Sym^2(H_2)$ has just a single trivial representation and we get no new lattice tensors relative to
the continuum.

At $a^{-2}$ we will again need $Sym^2(H_2) \otimes ( 1 \oplus H_2 )$ so let us fully decompose, 
\bea
Sym^2(H_2) &=& 2 \tau^{(1)}_1 \oplus \tau^{(1)}_2 \oplus 2 \tau^{(2)}_1 \oplus \tau^{(2)}_2
\oplus 2 \tau^{(3)}_1 \oplus \tau^{(6)}_1 \oplus 2 \tau^{(6)}_2
\oplus 2 \tau^{(6)}_3,
\eea
with the two trivial representations seen explicitly, but this time we further need to 
tensor the above with $( 1 \oplus \tau^{(3)}_1 \oplus \tau^{(6)}_3 )$, leading to the trivial
representation appearing a total of six times, which is 4 more than in the continuum. The six tensors
come about as $p^2$ times the two $a^{-4}$ tensors (one continuum, one lattice), the one continuum
$a^{-2}$ tensor, and three genuinely new $a^{-2}$ lattice tensors.

The same line of reasoning on the 16-cell honeycomb gives,
\bea
Sym^2(H_2) &=& \varrho^{(1)}_1 \oplus \varrho^{(2)}_1 \oplus \varrho^{(2)}_3 \oplus
\varrho^{(4)}_5 \oplus \varrho^{(6)}_2 \oplus 2 \varrho^{(9)}_1
\oplus \varrho^{(12)}_1 
\eea
which, if tensored with $1 \oplus \varrho^{(9)}_1$, gives the trivial three times, once from $1 \otimes 1$ and
two from $\varrho^{(9)}_1 \otimes \varrho^{(9)}_1$. The three tensors are the two known from the continuum and
only one new lattice tensor.

The counting rules can be understood also from the different orders of the independent 
invariant polynomials on the two lattices. On the cubic lattice there are independent invariant
polynomials of order $2,4,6,8$ while on the 16-cell honeycomb $2,6,8,12$. At second order we have of
course $p^2$, the same as in the continuum. On the cubic lattice the other three are,
\bea
p^{(2n)} = \sum_\mu p_\mu^{2n}  \qquad \qquad    n = 2,3,4,
\eea
while on the 16-cell honeycomb we have,
\bea
I_6 &=& 5 p^2 p^{(4)} - 4 p^{(6)} \nn \\
\label{poly}
I_8 &=& 35 \left(p^{(4)}\right)^2 + 28 p^2 p^{(6)} - 60 p^{(8)} \\
I_{12} &=& 22 p^2 p^{(10)} + 165 p^{(4)} p^{(8)} + 154 \left( p^{(6)} \right)^2 - 340 p^{(12)}. \nn
\eea
At $a^{-4}$ the new cubic tensor 
corresponds to $\partial_\mu \partial_\nu \partial_\rho \partial_\sigma p^{(4)} - (traces)$. 
The three genuinely new $a^{-2}$ cubic lattice tensors correspond to
\bea
\partial_\mu \partial_\rho\, p^{(4)} \partial_\nu \partial_\sigma\, p^2, \qquad
\partial_\mu \partial_\nu \partial_\rho\, p^{(4)} \partial_\sigma\, p^2, \qquad
\partial_\mu \partial_\nu \partial_\rho \partial_\sigma\, p^{(6)},
\eea
after again subtracting traces and made symmetric as the energy-momentum tensor 2-point function. 
On the 16-cell honeycomb there is no independent $4^{th}$
order polynomial, so no new tensor at $a^{-4}$ at all. At $a^{-2}$ the single available new lattice tensor
corresponds to $\partial_\mu \partial_\nu \partial_\rho \partial_\sigma \, I_6 - (traces)$, matching the numbers
perfectly we have obtained above from the irreducible representations.

Hence, overall, the number of new invariant power divergent lattice contact terms, not present in the continuum is
five on the cubic lattice and is only one on the 16-cell honeycomb, a considerable improvement.

Without details we simply state for completeness that a completely analogous classification 
at $a^0$ i.e. logarithmic terms, lead to the following reduction in the number of tensors. One needs an
invariant 8-index tensor contracted with $p_\mu p_\nu p_\rho p_\sigma$ and the possible number of
terms in the continuum is 3, on the cubic lattice there are 15 additional $O(4)$-breaking ones, however on the 
16-cell honeycomb there are only 5 additional $O(4)$-breaking ones.

\section{Conclusion}

Studying lattice QCD on the 16-cell honeycomb is a very promising approach. At fixed physical volume
there are twice as many sites as on the cubic lattice and 3 times as many nearest neighbors, leading
naively to a six times larger cost, however as we have seen in \cite{Katz:2025ycr} and \cite{Katz:2026qmx}, the
quantitative and qualitative improvements more than compensate for this increase. The reason is the
enlarged space-time symmetry relative to the cubic lattice, leading to smaller cut-off effects, less
mixing, more irreducibles for on-shell matrix elements at rest and similarly larger irreducibles
requiring fewer renormalization constants.

The general representation theoretic considerations translate into concrete gains for the energy-momentum
tensor, which in the continuum requires two renormalization constants, one for the gluonic and one for
the fermionic part. On the cubic lattice each splits up into two lattice symmetry irreducibles,
while on the 16-cell honeycomb each stays irreducible, leading to the same renormalization structure as
in the continuum. 

In this contribution we also reviewed the fate of doubler modes of the naive fermion discretization and
have seen that even though 44 out of the 71 doublers do break Lorentz (Euclidean) symmetry, the Wilson
term shifts all doublers to $\sim 1/a$ masses leaving behind the physical mode only, which was Lorentz
(Euclidean) invariant to begin with. 

\section*{Acknowledgements}

This work was supported by NKFIH grants TKP2021-NKTA-64, 
2026-4.1.1-MISSZIÓK-2026-00005, NKKP Excellence No. 151482 and K-147396.

\end{document}